\documentclass[a4paper, 12pt]{article}

\usepackage{tikz, floatrow, pgfplots, tkz-euclide} 
\usetikzlibrary{positioning, matrix, calc, patterns, plotmarks, shapes}
\usepgfplotslibrary{fillbetween, external}
\pgfplotsset{width=10cm, compat=1.9}
\usepackage{graphicx,soul, color}      
\usepackage[font=small,labelfont=bf]{caption} 
\usepackage{subfig}
\usepackage{multirow,array}

\usepackage[T1]{fontenc}
\usepackage{fourier}	
\usepackage{amsmath, mathtools, nicefrac}    
\usepackage{amsthm}		
\usepackage{enumitem}   
\usepackage{amsfonts}	
\usepackage{bm}	        
\usepackage[hang,flushmargin]{footmisc} 
\usepackage[authoryear, round]{natbib}
\usepackage{setspace,geometry}
\usepackage{enumitem} 
\usepackage{authblk}	
\usepackage{multirow,array}     

\DeclareMathSizes{11.5}{20}{14}{10}   
\renewcommand{\arraystretch}{1.2} 
\newtheorem*{define}{Definition}
\newtheorem{propos}{Proposition}

\newtheorem{assume}{Assumption}
\newtheorem{result}{Result}
\newtheorem{lemm}{Lemma}

\DeclareMathOperator{\E}{\mathbb{E}}
\DeclareMathOperator*{\R}{\mathbb{R}}
\DeclareMathOperator*{\Rf}{\mathcal{R}}
\DeclareMathOperator*{\Lo}{\mathcal{L}}

\DeclareMathOperator*{\T}{\mathcal{T}}
\DeclareMathOperator*{\Ind}{\mathbb{1}}

\DeclareMathOperator\supp{supp}

\usepackage{hyperref}
\hypersetup{colorlinks=false,pdfborder=0 0 0,citebordercolor={0 0 0}}

\begin{document}

\title{	
\normalfont \normalsize
\Large Evolution of Risk-Taking Behaviour and Status Preferences in Anti-Coordination Games\\[15pt] 
}

\author[1]{Manuel Staab}
\affil[1]{Aix-Marseille University (Aix-Marseille School of Economics), 5-9 Boulevard Maurice Bourdet, 13205 Marseille, France, email: manuelstaab@gmail.com}

\maketitle
\thispagestyle{empty}

\begin{abstract}
This paper analyses how risk-taking behaviour and preferences over consumption rank can emerge as a neutrally stable equilibrium when individuals face an anti-coordination task. If in an otherwise homogeneous society information about relative consumption becomes available, this cannot be ignored. Despite concavity in the objective function, stable types must be willing to accept risky gambles to differentiate themselves, and thus allow for coordination. Relative consumption acts as a form of costly communication. This suggests status preferences to be salient in settings where miscoordination is particularly costly.
 \\[25pt] 

\end{abstract}

\newpage

\section{Introduction}

In many economic settings, people's preferences seem to not only depend on their own allocation but also the allocations of the people around them.%
    \footnote{This goes back to \citet{veblenLeisure} and the concept of conspicuous consumption. See, for instance, \citet{frank85} and \citet{masmoretti09} for empirical examples for the economic importance of positional concerns, as well as (among many others) \citet{corneo1998}, \citet{hopkins04}, and \citet{becker05} for theoretical treatments.}
In particular, these can take the form of status or rank-dependent preferences as in \citet{robson92} and \citet{robray12}. Explanations for such preferences suggested in the literature include the allocation of goods outside market mechanisms, as described in \citet{coleEt}, as well as the information relative consumption might contain about the state of the environment, as in \citet{samuelson04}. While also employing an evolutionary approach, this paper instead focuses on their role for coordination. It is shown how even in a perfectly equal society, individuals might have a strong incentive to induce variation in consumption by taking risky gambles if this facilitates social interactions and avoids costly conflict. Furthermore, if information on relative consumption is available, it cannot be ignored by a stable type. This can lead to preferences over consumption rank. These insights imply more generally that if there is no costless way of coordinating in a society, it is evolutionary optimal for agents to engage in risky behaviour to generate signals, which then must be heeded by the other agents.

The main idea explored here is that while successful social interactions require a degree of cooperation, the roles people (or animals) need to take might be asymmetric; not everybody can be a leader at the same time. This is modelled as a 2-player anti-coordination game equivalent to a Hawk-Dove game, where a Dove player would rather face a Hawk than a Dove opponent. The focus lies on a one-population scenario where players are ex-ante identical and it is thus challenging for players to coordinate their actions. If individuals successfully coordinate by choosing distinct actions, they receive a positive but unequal payoff. If (anti-) coordination fails, the interaction is not productive. In other words, when players are unable to fill the roles necessary for the interaction, there might be costly conflict which lowers payoffs. In a biological context, this does not necessarily have to be conflict that causes the risk of injury but can also be the case of a ritualized display that requires effort as described in \citet{smith74} or it might simply be that the task, for example a joint hunt, fails. In the absence of any asymmetries, players randomize over actions. This is inefficient as outcomes where players choose the same action (e.g. both act hawkish or dovish) occur with positive probability. As discussed in \citet{smith1976logic}, if players have some type of `cue' available as to who takes which role, conflict can be avoided. While previous work analyzes similar settings with exogenously given cues and cheap-talk communication (e.g., \citet{herold20} and \citet{hurkens2003}), it is shown here how such cues emerge endogenously in the form of costly communication. For example, \citet{herold20} show how a set of freely available labels in a society can alleviate conflict in an anti-coordination game. The paper at hand, in contrast, shows how players themselves can create such labels and in fact will do so in an evolutionary stable state.

In the model examined here, players are endowed with an identical (subsistence) consumption allocation. They engage in an anti-coordination task with other players. This captures various kinds of social interactions where the roles people need to take are somewhat asymmetric. For instance, when several members of a group go hunting together, not everybody can determine where and what to hunt. While these situations are admittedly vastly more complex in the real world, it can be argued that this bilateral anti-coordination game still captures some relevant aspects. The outcomes of these interactions determine the players' allocations of a second good that will simply be called `social good'. This can be interpreted as access to socially awarded prizes, like mating or feeding privileges, or simply a different consumption good only attainable through cooperation. Both the consumption and social good determine a player's fitness and thus their reproductive success. Maximizing their fitness is the evolutionary objective of each player.

Before this interaction takes place, players have the possibility to accept gambles over their consumption endowment. For instance, a hunter-gatherer could either forage in a well-known area and obtain a relatively certain payoff, or explore new places with uncertain returns. Or an individual could engage in risky investment activities to obtain greater wealth than their peers. These gambles are taken to be not inherently beneficial (individual fitness is assumed to be concave in the consumption good), but as will be shown, they can create an observable heterogeneity in the population. This heterogeneity can then take the role of a message for the coordination problem, not unlike the signals analyzed in coordination games with cheap talk, such as in \citet{banerjee2000neutrally}. As a key distinction, messages here are verifiable and costly to generate. 

We are mostly interested in analysing stable preferences over lotteries and hence employ an `evolution of preferences' approach as in \citet{alger2013} and \citet{dekel2007evolution}. It is shown that the information provided by relative consumption cannot be ignored by a stable type. Preferences over outcomes in the anti-coordination game must be such that they only allow for equilibria that perfectly correlate actions whenever there are non-identical cues or messages available.%
    \footnote{See \citet{Smith79royal} for various examples where size determines outcomes in asymmetric interactions even though size is not necessarily correlated with the success in conflict. For instance, \citet{riechert1978} presents a particular example where spiders react to relative size as a cue in bilateral interactions. In the context of the model presented here, differences in consumption could lead to observable differences which can then be used to correlate actions.} 
When risky gambles are available, a `norm' emerges where consumption determines the roles in the anti-coordination game and individuals are willing to accept such risky gambles in equilibrium. This can lead to a `hierarchical' society (e.g., higher relative consumption leads to a larger allocation of the social good), but also to equilibria that are more egalitarian. It is shown further that stable choice behaviour over gambles can be represented by rank-dependent utility as in \citet{robray12}. The paper thus presents an evolutionary origin for rank-dependent preferences and their effect on risk-taking behaviour and competition.

Interpreted loosely, the consumption lotteries could also be a series of antagonistic interactions whose probabilistic outcomes determine access to resources but avoid more serious conflict, as observed in various animal societies. For instance, many primates establish a social hierarchy through (potentially costly) dominance behaviour that then determines access to resources like mating partners.%
    \footnote{See, for example, \citet{hausfater75dominance} for a discussion of hierarchies in male baboons and \citet{Abbott1991} for hierarchies in female marmosets.}  
As the outcome of these interactions can entail a degree of randomness and risk of injury, especially for animals of comparable strength, this can be considered a costly lottery. But this avoids more frequent and serious or even lethal conflict when it comes to accessing resources. In other words, a clearly established hierarchy avoids conflict when resources become available. More closely related, the model captures features of observed risk-taking behaviour in humans, especially in young male adults, who engage in high-stakes, high-risk activities to obtain social status and thus increase their chances of reproductive success.%
    \footnote{See, for instance, \citet{ellis12} for a detailed analysis of risk-taking in young adults. \citet{bottan2022} provide evidence for the importance of relative income when choosing where to live, particularly for single individuals.}

The remaining paper is structured as follows: Section \ref{SECTION:Model} formalizes the model, which is then analyzed in Section \ref{SECTION:results}. The analysis first focuses on the anti-coordination game itself (the `stage game') and then the meta game, where messages are generated. We demonstrate the existence of a stable type using a utility specification that takes into account relative consumption rank. Section \ref{SECTION:discuss} provides a brief discussion of the results. All omitted proofs are in Appendix \ref{appendix:proofs}.

\section{The Model}\label{SECTION:Model}

The population consists of a continuum of individuals of measure 1. Their evolutionary success is determined by the fitness function $f$, which is a mapping
\begin{equation*}
	f: {\R}^2_{+} \to {\R}.
\end{equation*}
The inputs, represented by the pair $(c,s) \in {\R}^2_+$, are taken to be an individual's resources. The first element ($c$) is interpreted as a \textit{consumption good} and the second element ($s$) as a socially obtained good or \textit{social good}. Individuals are initially endowed with a subsistence amount $c_0 > 0$ of the consumption good and no social good. They engage in a binary interaction through which, if successful, they gain $s$. This is interpreted as obtaining socially awarded rewards like influence, or mating partners. Both consumption and social good are assumed to be beneficial for evolutionary success but exhibit diminishing marginal returns. Furthermore, the consumption good is assumed to be essential for survival. This motivates the following assumptions on the fitness function $f$:

\begin{assume}\label{AssuCont}
	$f$ is strictly increasing in $c$ and $s$. It is continuous and
	at least twice differentiable in $c$ and $s$. 
\end{assume}

\begin{assume}\label{AssuConcav}
 $f$ is strictly concave in $c$ and $s$. Furthermore, $\lim_{c \to 0}f(c, s ) \to -\infty$, $\lim_{c \to 0} \frac{\partial f(c, s )}{\partial c} \to \infty$, and $\lim_{c \to \infty} \frac{\partial f(c, s)}{\partial c} \to 0$ for every $s \in {\R}_+$.
\end{assume}

Individuals have access to a set of consumption \textit{lotteries} $\Lo$. If they choose a lottery $L \in \Lo$, they obtain a new consumption endowment $c_0 + C$, where $C$ is a random variable that follows a probability distribution as defined by $L$. The realizations of a given lottery across individuals are assumed to be statistically independent. For notational simplicity, we identify a lottery by its distribution function, i.e., $L(x)$ gives the probability that $L$ yields an outcome of at most $x$. The corresponding probability measure is denoted by $\lambda_L$. We restrict attention to Borel-measurable distributions. We call a lottery \textit{trivial} if its support is a singleton. The trivial lottery with support $\{0\}$ is assumed to be always available in $\Lo$. Lotteries are assumed to be bounded, meaning $\supp(L)$ is contained in some finite interval. As $f$ is defined over $\R_+^2$, we assume that an individual with consumption $c$ has only access to gambles with the support bounded below by $-c$. Furthermore, lotteries are assumed to be fair, i.e., $E_{L}\bigl[C\bigr] = 0$ meaning there is no inherent gain or loss in accepting a gamble. The analysis readily extends to cases where gambles are costly.

\begin{define}[Lottery]\label{define:Lottery}
	A fair and bounded consumption lottery $L \in \Lo$ is a Borel measurable probability distribution over ${\R}$ with its support contained in some bounded interval $V \subset \R$ and $\int_V x dL(x) = 0$.
\end{define}

\noindent\textbf{Timing.} First, individuals choose a consumption lottery (meta game). After the lottery outcome is realized, they engage in a binary anti-coordination game that yields social good $s$ (stage game). They then achieve reproductive success according to the fitness function $f(c,s)$.\\
\\
\noindent\textbf{Stage game.} Agents take part in a symmetric, two-player anti-coordination game which determines their allocation of $s$. This is referred throughout as the stage game. For each player, this is modelled as a single interaction with an opponent whose strategy is representative of the strategies in the population. More specifically, if some part of the population plays one strategy while the remaining fraction plays another, then this representative opponent follows a strategy that is the mixture of both with weights equal to the frequency in the population. In line with the standard interpretation in the literature (e.g. \cite{weibull97}), this is seen as analogous to individuals interacting frequently and randomly with others. The expected payoff in the anti-coordination game then stands for the social good that an individual receives on average from interactions with other members of society.

Individuals can either take the leading role and act \textit{hawkish} (h) or defer and act \textit{dovish} (d). Players are anonymous meaning the game has no (ex-ante) uncorrelated asymmetries in the sense of \citet{smith1976logic}. The general payoff structure is summarized in Table \ref{tab:payoff_matrix}. As we are interested in anti-coordination games where players prefer to take unequal actions, it is required that the payoffs on the off-diagonal are greater than the payoffs on the diagonal: $s_2 > \max\{s_1, s_4\}$ and $s_3> \max\{s_1,s_4\}$.

\begingroup
\renewcommand{\arraystretch}{1.5} 
\begin{table}[hbt!]
    \subfloat[general]{\begin{tabular}{r | c | c |}
        \multicolumn{1}{c}{} & \multicolumn{1}{c}{$h$}  & \multicolumn{1}{c}{$d$} \\    \cline{2-3}
        $h$   & $s_1$   & $s_2$\\ \cline{2-3}
        $d$   & $s_3$   & $s_4$\\ \cline{2-3}
    \end{tabular}}
    \quad \quad \quad \quad
    \subfloat[simplified]{\begin{tabular}{r | c | c |}
        \multicolumn{1}{c}{} & \multicolumn{1}{c}{$h$}  & \multicolumn{1}{c}{$d$} \\    \cline{2-3}
        $h$   & $0$   & $\overline{s}$\\ \cline{2-3}
        $d$   & $\underline{s}$   & $0$\\ \cline{2-3}
    \end{tabular}}
    \caption{Payoff matrices of a stage game with the anti-coordination property.}
    \label{tab:payoff_matrix}
\end{table}
\endgroup
For expositional clarity, payoffs from Table \ref{tab:payoff_matrix} will be normalized such that $s_1 = s_4 = 0$, while $\overline{s} \coloneqq s_2$, $\underline{s} \coloneqq s_3$, with the assumption that $\overline{s} > \underline{s}$. The results, however, go through with the general case of $s_2 > s_3$ and $s_3 > \max\{s_1, s_4\}$. The payoff of a player $i$ choosing action $h$ with probability $\phi_i$ against an opponent $j$ choosing $h$ with probability $\phi_j$ can be summarized as:
$$
    S(\phi_i, \phi_j) \coloneqq \phi_i \, (1-\phi_j) \, \overline{s} \; + \; (1-\phi_i) \, \phi_j \, \underline{s}.
$$

This game has the `anti-coordination property' as defined in \citet{kojima2007anti}, meaning that the support of any mixed strategy includes all the worst responses. The only evolutionary stable strategy is then the (unique) interior Nash equilibrium, i.e., $h$ is played with probability $\frac{\underline{s}}{\underline{s}+\overline{s}}$. However, this is inefficient as aggregate payoffs are not on the Pareto-frontier. The average payoff is less than the efficient $\frac{\overline{s}+\underline{s}}{2}$ and, in fact, less than $\underline{s}$.\\
\\
\noindent\textbf{Meta game.} Before the stage game is played, both players can observe their opponent's consumption level $c$. Similar to (anti-)coordination games with a cheap talk stage, this acts as a message or label that can help to coordinate players' actions. However, because of the concavity of $f$ in $c$, inducing variation in consumption through lotteries is costly. 
The choice of lottery before the stage game is played can be interpreted as a meta game, where each player decides on a probability distributions over messages they send. The message space $M \subset \R^+$ is the set of possible consumption levels that can be reached with any lottery. More precisely, 
\begin{equation*}
    M \coloneqq \bigl\{m \in {\R}_+ \; | \; m = c_o + C, \; \exists L \in \Lo \text{ s.t. } C \in \supp(L)   \bigr\}.
\end{equation*}

Messages and consumption levels are treated interchangeably, meaning $m$ and $c$ are used synonymously. Given an aggregate distribution over consumption levels $G$ in the population, the `effective' message space, i.e., the set of messages that can arise from the lotteries actually chosen, is denoted by:
\begin{equation*}
    M_G \coloneqq \bigl\{m \in {\R}_+ \; | \; m \in \supp(G) \bigr\}.
\end{equation*}

After observing their opponent's message (i.e., consumption level), a player decides which action to choose in response. This is characterized by the function $\rho: M \to A$. The set of all such responses is denoted by $\Rf$, with the set of strategies equal to the set of all probability distributions over such responses $\Phi \coloneqq \Delta(\Rf)$. Naturally, in the absence of variation in the consumption distribution, this is identical to the set-up without a messaging stage.

The strategy space in the meta game, which includes the lottery choice from $\Lo$, is denoted by $\Theta \coloneqq  \Delta(\Lo) \times \Phi^M$, where $\Phi^M$ contains the possible strategies in the stage game for each possible $m \in M$ and $\Delta(\Lo)$ is the mixture space of available lotteries.%
    \footnote{See \citet{kreps1988}, Definition 5.10. If $\Lo$ contains all fair lotteries with their support bounded by some interval $V \subset \R$, we can just replace $\Delta(\Lo)$ by $\Lo$, which is closed under convex combinations.} 
For any element $\theta \in \Theta$, we denote by  $\lambda_{\theta}(m-c_0)$ the probability that a player with strategy $\theta \in \Theta$ sends a message $m$, where $L_{\theta} \in \Delta(\Lo)$ is the (compound) lottery corresponding to the choice in $\Delta(\Lo)$ prescribed by $\theta$. The strategies in the stage game are assumed to be measurable with respect to the message space. We denote by $\phi_{\theta}(m' | m)$, the (conditional) probability that a player with strategy $\theta$, who sent a message $m$ and observes $m'$ from their opponent, adopts a response $\rho(m')=h$.\\

\noindent\textbf{Types.} A type $\tau$ describes a utility function over outcomes $u_{\tau}: \Theta \times \Theta \to \R$. The space of all such types is denoted by $\T$. If there is no multiplicity, a strategy chosen by a type $\tau$ in equilibrium is denoted by $\theta_{\tau} \in \Theta$, with the corresponding lottery and stage game strategy referred to as $L_{\tau}$ and $\phi_{\tau}$.\\

\noindent\textbf{Equilibrium \& Evolutionary stability.}
We closely follow the indirect evolutionary approach of \citet{alger2013}. A \textit{population configuration} describes the set of types and their relative frequency in the population of players. As we are interested in (evolutionary) stable outcomes, we focus on (binary) configurations with a resident and a mutant type. A triple $(\tau, \tau', \epsilon)$ then describes a population with an resident type $\tau$,who has mass $(1-\epsilon)$, and a mutant type $\tau'$ with mass $\epsilon$. If $\epsilon=0$, we refer to it as a pure population. Given a population configuration, in any Nash equilibrium both types are best-responding to a representative opponent, where the best-response depends on their preferences. In other words, types are not observable.

\begin{define}[Nash equilibrium]
Given a population configuration $(\tau, \tau', \epsilon)$, a pair $(\theta, \theta') \in \Theta \times \Theta$, is a Nash-equilibrium of the meta game if 
\begin{equation*}
\begin{split}
    &\theta \in \arg\max_{\theta \in \Theta} \; (1-\epsilon) \cdot u_{\tau}(\theta, \theta) + \epsilon \cdot u_{\tau'}(\theta, \theta'),\\
    &\theta' \in \arg\max_{\theta' \in \Theta} \; (1-\epsilon) \cdot u_{\tau'}(\theta', \theta) + \epsilon \cdot u_{\tau'}(\theta', \theta').
\end{split}
\end{equation*}
\end{define}

The overall fitness payoff depends on the choice of lottery as the response to the message received. If the population configuration is $(\tau, \tau', \epsilon)$, with a resident playing strategy $\theta_{\tau}$ and a mutant $\theta_{\tau'}$, then the the expected fitness outcome of a resident type can be written as:
\begin{equation*}
    \begin{split}
            V_{\tau}(\tau, \tau', \epsilon) \coloneqq  (1-\epsilon)\cdot & \E_{m}\Bigl[ \E_s\bigl[f(m,s) \; | \; \phi_{\tau}(m'|m), \phi_{\tau}(m|m')\bigr] \; \big|\; L_{\tau}, L_{\tau}\Bigr]\\
        \epsilon \cdot & \E_m\Bigl[ \E_s\bigl[f(m,s) \; | \; \phi_{\tau}(m'|m), \phi_{\tau'}(m|m')\bigr] \; \big| \; L_{\tau} , L_{\tau'}\Bigr]
    \end{split}
\end{equation*}
where $L_{\tau}$ is the lottery choice of a type $\tau$, and $\phi_{\tau}$ the corresponding strategy in the stage game. Similarly, the outcome of a mutant is denoted by $V_{\tau'}(\tau, \tau', \epsilon)$.

\begin{define}[Evolutionary Stability]
A type $\tau \in \T$ is \textbf{neutrally stable} against a type $\tau' \ne \tau$ if there exists an \,$\overline{\epsilon} > 0$ such that for all \,$\epsilon < \overline{\epsilon}$ and all Nash equilibria of the meta game:
\begin{equation*}
    V_{\tau}(\tau, \tau', \epsilon) \ge V_{\tau'}(\tau, \tau', \epsilon).
\end{equation*}
A type $\tau$ is \textbf{neutrally stable} if this holds for all $\tau' \ne \tau$, $\tau' \in \T$. It is \textbf{evolutionary stable} if the inequalities are strict.
\end{define}

As neutral stability is necessary for evolutionary stability, in results pertaining to both we simply refer to the \textit{stability} of a type.

\section{Results}\label{SECTION:results}

The analysis focuses on identifying the choice behaviour over lotteries that, together with the strategy in the stage game, render a type (evolutionary) stable. We first focus on stable strategies in the stage game conditional on a consumption distribution. We then proceed by analysing the lottery choices that determine this distribution. We characterise stable preferences over lotteries in the meta game and demonstrate how positional concerns can arise in such a setting.

\subsection{Stability \& the stage game}\label{section:stableStageGame}

This section identifies stable preferences over outcomes in the stage game and the strategies they induce in equilibrium. The main take-away is that if players observe different messages in an interaction, then stability requires that in \textit{any} equilibrium they use this information to achieve a Pareto-efficient allocation of the social good. Note that here efficiency simply requires an average allocation of $\frac{\overline{s}+\underline{s}}{2}$ across the two players, meaning they never miscoordinate. This does, however, not imply that this allocation maximizes fitness. A type with preferences that lead to a Pareto-efficient allocation can only be invaded if the aggregate allocation allows for a superior lottery choice. Only differences in risk-taking behaviour can replace efficient but overall suboptimal strategies in the stage game. 

The first part of this statement is established in Lemma \ref{lemm:coordActionsStage}. If two players of a stable type meet and their consumption levels differ, then those fully determine their action choices. Observable, non-identical messages must remove any stochasticity in choices and lead to a Pareto-efficient social good allocation in their interaction. 

\begin{lemm}\label{lemm:coordActionsStage}
    A type $\tau \in \T$ is stable only if there exists an $\; \overline{\epsilon}>0$ such that for any equilibrium strategy $\; \theta_{\tau} \in \Theta$ given a configuration $(\tau, \tau', \epsilon)$ with $\;\epsilon < \overline{\epsilon}$, and almost all distinct messages $m, m' \in M$, we have $\phi_{\tau}(m'|m) \in \{0,1\}\;$ and $\; \phi_{\tau}(m|m') = 1-\phi_{\tau}(m'|m)$.
\end{lemm}

Table \ref{tab:payoff_matrix_unstable} shows an example for preferences over outcomes in the stage game that allow for an interior equilibrium. As follows from Lemma \ref{lemm:coordActionsStage}, a type whose preferences can be represented by these payoffs cannot be stable. A mutant type that perfectly correlates actions based on messages can achieve a strictly superior outcome in this equilibrium, thus violating the requirement for stability. Table \ref{tab:payoff_matrix_rel} demonstrates an alternative set of preferences that yield a unique equilibrium in the stage game whenever messages are distinct. Any pairing of players with distinct messages then results in an efficient allocation of $s$. If in every equilibrium such a type chooses a continuous distribution over messages, i.e., an atomless consumption lottery, then identical messages and consequently miscoordination among these players becomes a probability $0$ event. Such a type is evolutionary stable against any other type that has the same preferences over lotteries (Lemma \ref{lemm:stageGameStable}).

\begingroup
\renewcommand{\arraystretch}{1.5} 
\begin{table}[hbt!]
    \subfloat[$m > m'$]{\begin{tabular}{r | c | c |}
        \multicolumn{1}{c}{} & \multicolumn{1}{c}{$h$}  & \multicolumn{1}{c}{$d$} \\    \cline{2-3}
        $h$   & $0$   & $\overline{u}$\\ \cline{2-3}
        $d$   & $\underline{u}$   & $0$\\ \cline{2-3}
    \end{tabular}}
    \quad \quad \quad \quad
    \subfloat[$m < m'$]{\begin{tabular}{r | c | c |}
        \multicolumn{1}{c}{} & \multicolumn{1}{c}{$h$}  & \multicolumn{1}{c}{$d$} \\    \cline{2-3}
        $h$   & $0$   & $\overline{u}$\\ \cline{2-3}
        $d$   & $\underline{u}$   & $0$\\ \cline{2-3}
    \end{tabular}}
    \quad \quad \quad \quad
    \subfloat[$m = m'$]{\begin{tabular}{r | c | c |}
        \multicolumn{1}{c}{} & \multicolumn{1}{c}{$h$}  & \multicolumn{1}{c}{$d$} \\    \cline{2-3}
        $h$   & $0$   & $\overline{u}$\\ \cline{2-3}
        $d$   & $\underline{u}$   & $0$\\ \cline{2-3}
    \end{tabular}}
    \caption{Example for preferences over stage game outcomes that allow for an inefficient interior equilibrium. A type with such preferences cannot be evolutionary stable.}
    \label{tab:payoff_matrix_unstable}
\end{table}
\endgroup

\begingroup
\renewcommand{\arraystretch}{1.5} 
\begin{table}[hbt!]
    \subfloat[$m > m'$]{\begin{tabular}{r | c | c |}
        \multicolumn{1}{c}{} & \multicolumn{1}{c}{$h$}  & \multicolumn{1}{c}{$d$} \\    \cline{2-3}
        $h$   & $\underline{u}$   & $\overline{u}$\\ \cline{2-3}
        $d$   & $0$   & $0$\\ \cline{2-3}
    \end{tabular}}
    \quad \quad \quad \quad
    \subfloat[$m < m'$]{\begin{tabular}{r | c | c |}
        \multicolumn{1}{c}{} & \multicolumn{1}{c}{$h$}  & \multicolumn{1}{c}{$d$} \\    \cline{2-3}
        $h$   & $0$   & $0$\\ \cline{2-3}
        $d$   & $\overline{u}$   & $\underline{u}$\\ \cline{2-3}
    \end{tabular}}
    \quad \quad \quad \quad
    \subfloat[$m = m'$]{\begin{tabular}{r | c | c |}
        \multicolumn{1}{c}{} & \multicolumn{1}{c}{$h$}  & \multicolumn{1}{c}{$d$} \\    \cline{2-3}
        $h$   & $0$   & $\overline{u}$\\ \cline{2-3}
        $d$   & $\underline{u}$   & $0$\\ \cline{2-3}
    \end{tabular}}
    \caption{Example for preferences over stage game outcomes that yield a unique equilibrium in which actions are perfectly correlated if $m' \ne m$, as is necessary for stability.}
    \label{tab:payoff_matrix_rel}
\end{table}
\endgroup

\begin{lemm}\label{lemm:stageGameStable}
    Suppose there exists a set of types $\hat{\T} \subset \T$ and an $\; \overline{\epsilon}>0$ such that for all distinct $\tau, \tau' \in \hat{T}$ and any configuration $(\tau, \tau', \epsilon)$ with $\epsilon < \overline{\epsilon}$, both types choose an identical continuous lottery in any equilibrium. Then a type $\tau \in \hat{\T}$ is neutrally stable against any $\tau' \in \hat{\T}$ if and only if in any equilibrium $\phi_{\tau}(m'|m) \in \{0,1\}$ and $\phi_{\tau}(m|m') = 1- \phi_{\tau}(m'|m)$ for almost all $m, m' \in M$. 
\end{lemm}

A strategy in the stage game that perfectly correlates which player is to lead (play $h$) and which player is to follow (play $d$) efficiently allocates the social good. Similar to a correlation device in the sense of \citet{aumann74}, distinct messages allow for such an efficient outcome. The difference here is that there must be no equilibrium that fails to make use of the available information. Preferences need to adjust to achieve a unique outcome. This outcome, however, might still be suboptimal. Even if the norm that determines who is to lead (and hence to receive the larger social good allocation) fails to maximize aggregate fitness for a given lottery choice, a successful invasion requires a mutant to choose a different lottery with a superior fitness outcome. If no such lottery exists, any Pareto-efficient norm is stable.

\subsection{Stability \& the meta game}

Stability requires that if distinct messages are available, these are used to fully coordinate actions. We now turn to the question if and how such messages arise in equilibrium. While any lottery itself is costless, strict concavity of $f$ in $c$ implies that choosing a risky gamble has in expectation a negative effect on fitness. Absent any other incentive, fitness maximization would require not accepting any (non-trivial) lottery. Lemma \ref{lemm:lotteryChoice} shows that despite the strict concavity of fitness in $c$, sending distinct messages through a consumption lottery is necessary for stability if suitable lotteries are available. In other words, independent of which strategy is used in the stage game to correlate actions,  we can always find a (fair) lottery (and in fact infinitely many lotteries) that generates signals at sufficiently low cost, such that the fitness benefit from coordination outweighs the fitness cost of the gamble. 

\begin{lemm}\label{lemm:lotteryChoice}
    There exists a set of lotteries $\Lo$ such that a type $\tau \in \T$ that chooses the trivial lottery in equilibrium cannot be stable.
\end{lemm}

The intuition is simple: if there is a mass point in the distribution of messages, coordination necessarily fails for a set of messages with positive measure. Such pairings lead to a Pareto-inefficient allocation of the social good. By taking on some arbitrarily small additional risk, a mutant can avoid this mass point and coordinate actions with other mutants (and possibly even the resident). This leads to a strictly higher fitness payoff in expectation. A mutant with preferences that favor such a modified lottery can successfully invade. Lemma \ref{lemm:lotteryChoice} thus allows us to conclude that if the set of lotteries $\Lo$ includes all fair lotteries bounded by some interval,%
    \footnote{Recall that since $f$ is only defined over $\R_+$, we assume an individual with consumption allocation $c$ has only access to lotteries that are bounded below by $c$.}
then a stable type must take on some risk and, in fact, choose a continuous lottery without mass points.

This raises the question what an equilibrium lottery choice `looks like' and if it even exists for a sufficiently rich $\Lo$. To address this, we derive a set of sufficient conditions for the stability of a type (Proposition \ref{propos:linearityLottery}). This implicitly characterizes an equilibrium lottery choice through the equilibrium distribution of fitness in a population. Ultimately, the question is whether we can find (neutrally) stable preferences over lotteries and stage game outcomes, and hence over strategies, that lead to the choice of such a lottery. This is the focus of Section \ref{section:statusPreferenes}.\\
\\
\noindent\textbf{Sufficient conditions for stability}\\
If a type $\tau$ is stable, then for any population configuration with a sufficiently small fraction of mutants, we must have $\phi_\tau(m'|m) \in \{0,1\}$ and $\phi_\tau(m|m') = 1-\phi_\tau(m'|m)$ (Lemma \ref{lemm:coordActionsStage}). If such types interact in the stage-game, they obtain a social good allocation of either $\underline{s}$ or $\overline{s}$. With a negligible fraction of mutants (or mutants that choose the same stage game strategy) and a continuous lottery choice $L$, the expected fitness of this type $\tau$ with message $m$ (i.e., a consumption outcome equal to $m$) can be written as:
\begin{equation*}
        F_{\tau}(m|L) \coloneqq \int \Bigl( \phi_\tau(x|m) \cdot f(m,\overline{s}) + \bigl(1-\phi_\tau(x|m)\bigr) \cdot f(m,\underline{s})\Bigr) \lambda_L(x-c_0) dx.
\end{equation*}
The probability that a player who sends message $m$ and receives $m'$ plays $h$ (i.e., $\phi_\tau(m'|m)$) takes the form of an indicator function on the support of $L$. The lottery choice in the population, however, might not be unique, particularly with mutants present. For a continuous aggregate consumption distribution $G$, this becomes:
\begin{equation}\label{eq:expFitnessGivenCons}
        F_{\tau}(m|G) \coloneqq \int \Bigl( \phi_\tau(x|m) \cdot f(m,\overline{s}) + \bigl(1-\phi_\tau(x|m)\bigr) \cdot f(m,\underline{s})\Bigr) d G(x).
\end{equation}
Maintaining the assumption that $\theta_{\tau}(m'|m) = \theta_{\tau'}(m'|m)$ for almost all $m,m' \in M$, the expected fitness of a type $\tau$, choosing a lottery $L$, can then be written as:
\begin{equation}\label{eq:expFitnessOverall}
        V_{\tau}(\tau,\tau',\epsilon) = \int \int \Bigl( \phi_\tau(x|m) \cdot f(m,\overline{s}) + \bigl(1-\phi_\tau(x|m)\bigr) \cdot f(m,\underline{s})\Bigr) \; dG(x) \; \lambda_L(m-c_0) dm.
\end{equation}

$F_{\tau}(m|G)$ plays a crucial role in establishing whether preferences can be stable. Independent of which exact utility function a type is optimizing, stability requires that there are no further gains from taking on risk at almost all consumption levels that are reached in equilibrium. If such gains exist, then it can be seen from \eqref{eq:expFitnessOverall} that there must be a lottery choice at $c_0$ that achieves higher expected fitness. A type with preferences that favour such a lottery could then achieve a strictly better fitness outcome. At the same time, taking on risk must be weakly beneficial for almost all messages, as otherwise a type could benefit by choosing a lottery that is less risky for a given set of messages. Consequently, if  $F_{\tau}(m|G)$ is linear over messages in $M_G$, both requirements are satisfied. This yields sufficient conditions for the neutral stability of a type (Proposition \ref{propos:linearityLottery}).

Let $V_L$ be the support of a lottery $L \in \Lo$, and $M_L(c) \coloneqq \bigl\{ x \in \R \; \big| \; (x-c) \in V_L\bigr\}$ the support of $L$ shifted by $c$, which corresponds to the set of possible messages an individual with endowment $c$ can send with lottery $L$.

\begin{propos}\label{propos:linearityLottery}
     A type $\tau \in \T$ is neutrally stable if for any distinct type $\tau' \in \T$ there exists an $\overline{\epsilon}>0$ such that for any configuration $(\tau, \tau', \epsilon)$ with $\epsilon < \overline{\epsilon}$ and corresponding equilibrium strategy $\theta_{\tau} \in \Theta$ with lottery choice $L$, the following hold:
     \begin{itemize}
         \item[(i)] $\phi_{\tau}(m'|m) \in \{0,1\}$, $\phi_{\tau}(m|m') = 1-\phi_{\tau}(m'|m)$ for almost all $m, m' \in M_G$,
         \item[(ii)] $\int F_{\tau}(m'|G) \lambda_{L'}(m'-m) dm'= F_{\tau}(m|G)$ for all $L'\in\Lo$ with $M_{L'}(m) \subseteq M_L(c_0)$, and almost all $m \in M_L(c_0)$,
         \item[(iii)] $\int F_{\tau}(m''|G) \lambda_{L''}(m''-m) d m''\leq F_{\tau}(m|G)$ for all $L'' \in \Lo$, and almost all $m \in M_L(c_0)$,
     \end{itemize}
    where $G$ is the aggregate distribution over messages and $L$ the lottery choice of type $\tau$.
\end{propos}

\subsection{Status preferences}\label{section:statusPreferenes}

While Lemma \ref{lemm:coordActionsStage} and Proposition \ref{propos:linearityLottery} present necessary conditions in the stage game and sufficient conditions in the meta game for a type to be neutrally stable, it remains to be shown that such preferences, and consequently such a type, indeed exist. We proceed by first identifying preferences over lotteries and stage game outcomes that are consistent with a stable type and then show how these can be combined into preferences over $\Theta$. More specifically, as discussed in Section \ref{section:stableStageGame}, we can transform payoffs in the stage game for different message combinations such that these yield strategies consistent with stability in any equilibrium. Preferences over strategies in the meta game (i.e., the lottery choice stage) can be captured by preferences over lotteries conditional on a consumption level and an aggregate consumption distribution. These can be represented by a utility function $V(L\;|\; c, G)$ that evaluates lotteries in $\Lo$ for an individual with consumption endowment $c$, conditional on some aggregate distribution $G$ over consumption (and hence messages). A lottery $L$ that maximizes $V(L\;|\;c, G)$ is a best response for an individual with message $c$ to an opponent sending messages according to $G$.%
    \footnote{This corresponds to an opponent with endowment $c_0$ that selects a lottery $L_G$, where $L_G$ is defined such that $L_G(x) \equiv G(c_0 + x)$, for all $x$ such that $c_0 + x$ is in the support of $G$.}
If we can find preferences over lotteries that in equilibrium yield strategies consistent with the conditions outlined in Proposition \ref{propos:linearityLottery}, then the corresponding type is neutrally stable. Proposition \ref{propos:statusPreferences} establishes this existence. The proof relies on a particularly salient utility specification: utility is determined by consumption as well as relative consumption rank and preferences in the stage game are such that individuals either strictly prefer to play $h$ or $d$, depending on whether they have higher or lower relative consumption than their opponent.

\begin{propos}\label{propos:statusPreferences}
There exist preferences over lotteries $V(L \; | \; c, G)$ and preferences over outcomes in the stage game such that a corresponding type is neutrally stable.
\end{propos}

While the complete proof can be found in the Appendix, we try to build some intuition here: Conditioning play in the stage game on relative consumption yields an efficient outcome whenever messages are distinct. Let $\phi^*$ denote such a strategy:
  \begin{equation}\label{eq:stageStratRelCon}
        \phi^*(m'|m) = 
        \begin{cases}
        1,    & \text{if } m > m'\\
        0,    & \text{if } m < m'\\
        \frac{f(m,\underline{s})}{f(m,\overline{s}) + f(m,\underline{s})},                  & \text{otherwise}.
    \end{cases}
    \end{equation}
For stability, this needs to be a unique equilibrium, which depends on the preferences over outcomes in the stage game. Fortunately, it can be easily verified that such preferences exist. Table \ref{tab:payoff_matrix_rel_stable} presents an example for payoffs that yield \eqref{eq:stageStratRelCon} as a unique Nash-equilibrium for any $\overline{u} > 0$, $u_1(m) = f(m,\overline{s})$, and $u_2(m) = f(m,\underline{s})$.%
    \footnote{Note that these are a slightly adjusted version of the payoffs shown in Table \ref{tab:payoff_matrix_rel}.}

\begingroup
\renewcommand{\arraystretch}{1.5} 
\begin{table}[hbt!]
    \subfloat[$m > m'$]{\begin{tabular}{r | c | c |}
        \multicolumn{1}{c}{} & \multicolumn{1}{c}{$h$}  & \multicolumn{1}{c}{$d$} \\    \cline{2-3}
        $h$   & $0$   & $\overline{u}$\\ \cline{2-3}
        $d$   & $0$   & $0$\\ \cline{2-3}
    \end{tabular}}
    \quad \quad \quad \quad
    \subfloat[$m < m'$]{\begin{tabular}{r | c | c |}
        \multicolumn{1}{c}{} & \multicolumn{1}{c}{$h$}  & \multicolumn{1}{c}{$d$} \\    \cline{2-3}
        $h$   & $0$   & $0$\\ \cline{2-3}
        $d$   & $\overline{u}$   & $0$\\ \cline{2-3}
    \end{tabular}}
    \quad \quad \quad \quad
    \subfloat[$m = m'$]{\begin{tabular}{r | c | c |}
        \multicolumn{1}{c}{} & \multicolumn{1}{c}{$h$}  & \multicolumn{1}{c}{$d$} \\    \cline{2-3}
        $h$   & $0$   & $u_1(m)$\\ \cline{2-3}
        $d$   & $u_2(m)$   & $0$\\ \cline{2-3}
    \end{tabular}}
    \caption{Preferences over stage game outcomes that yield $\phi^*$ as the unique equilibrium strategy.}
    \label{tab:payoff_matrix_rel_stable}
\end{table}
\endgroup

With a hierarchical strategy $\phi^*$ in the stage game, the expected fitness given some consumption distribution is determined by the measure of individuals with lower and higher consumption levels. Suppose $G$ is continuous,%
    \footnote{The proof in Appendix \ref{appendix:proofs} also treats the discontinuous case.}
then a player with message $m$ receives a stage game payoff of $\overline{s}$ against all players with $m'<m$. They interact with such a player with probability $G(m)$. The corresponding fitness payoff equals $f(m,\overline{s})$. Equivalently, with probability $1-G(m)$, they achieve a fitness outcome of $f(m,\underline{s})$. The expected fitness of an individual with consumption $m$ then equals:
    \begin{equation*}
        v(m,G(m)) \coloneqq G(m) \cdot f(m,\overline{s}) + \bigl(1-G(m)\bigr) \cdot f(m,\underline{s}).
    \end{equation*}
Stability requires an individual to choose a lottery that maximizes this in expectation. This defines preferences over lotteries, conditional on the population distribution $G$ and the endowment level $m$. For any $L \in \Lo$, these preferences can be represented as follows:
    \begin{equation*}
        V(L \; | \; m, G) = \int v\bigl(m+x, G(m+x)\bigr) \lambda_{L}(x-m) dx,
    \end{equation*}
where $\lambda_{L}$ is the probability measure corresponding to $L$. As is shown as part of the proof of Proposition \ref{propos:statusPreferences}, for such preferences there exists a unique equilibrium distribution $G$ over messages in a pure population.

As a final step, it remains to be shown that these preferences over lotteries and stage game outcomes can be combined to preferences over strategies in $\Theta$ that result in the same choices. This can be achieved by simply setting payoffs equal to $V(L| c, G)$ if strategies in the stage-game correspond to $\phi^*$, and $0$ otherwise. For a strategy $\theta'$, let $G_{\theta'}$ be the distribution over messages generated by an opponent choosing the lottery $L_{\theta'}$ (as prescribed by $\theta'$). Then the utility function $u_{\tau}$ over strategies can be defined as follows:
\begin{equation}\label{eq:stablePrefoverStrat}
    u_{\tau}(\theta, \theta') \coloneqq V(L_{\theta} \;| \; c_0, G_{\theta'}) \cdot {\Ind}_{\phi_{\theta} = \phi^*}.
\end{equation}
In a pure population, this yields an equilibrium message distribution $G$ that linearizes $V(L_{\theta} \;| \; c_0, G_{\theta'})$ over the support of $G$ and renders it strictly concave elsewhere. This is neutrally stable against any type that chooses a lottery with outcomes contained in the support of $G$, and is evolutionary stable against any other type.

\section{Discussion}\label{SECTION:discuss}

\noindent\textbf{Neutral stability \& evolution of preferences.} This analysis focuses primarily on neutral stability. Figure \ref{fig:relativeConsHierarchy} helps to illustrate why stronger notions cause existence problems in this setting: any non-trivial equilibrium lottery choice requires individuals to be indifferent between this and the trivial lottery. Over the relevant range, utility as well as expected fitness need to be linear. As shown in the graph, $v(c,r_G(c))$ takes the form of a straight line on the support of $G$ (shaded region). Any fair lottery with its support contained in this interval results in the same expected outcome. If a mutant chooses such a lottery, then for small $\epsilon$, $G$ remains an equilibrium distribution. The resident type chooses a suitably adjusted lottery that yields the same aggregate distribution. In this equilibrium, both types achieve the same expected fitness. A small fraction of more cautious risk-takers can `hide' in a population of more risky individuals. The resident type is neutrally stable. 

A mutant type who chooses riskier lotteries than such a neutrally stable resident, however, achieves strictly lower expected fitness. This follows from Proposition \ref{propos:linearityLottery} (iii). As $\epsilon \to 0$, the equilibrium distribution approaches $G$. Since $v(c,r_G(c))$ is strictly concave for sufficiently large gambles, resident types are not willing to choose more dispersed lotteries. A neutrally stable type is evolutionary stable against any mutant who takes on excessive risks. 

This also highlights the importance of focusing on the evolution of preferences over lotteries, rather than a fixed lottery choice. Any small fraction of mutants affect the equilibrium distribution of messages. No single lottery choice might be a best-response to all such mutant choices. Suitable preferences, however, can align a type's strategies with the evolutionary objective, while limiting equilibria to those in which this type achieves a (weakly) higher payoff than all mutants.\\
\\

\begin{figure}[ht!]
        \begin{tikzpicture}[scale=0.75]
        \definecolor{colorG}{RGB}{0,140,90}
        \definecolor{colorR}{RGB}{200,20,70}
            \begin{axis}[grid=none,
                ymin=-8,
                ymax=8,
                xmax=4.5,
                xmin=-0.01,
                samples=101,
                axis x line shift=-0,
                ytick={0},
                xticklabel={},
                yticklabel={},
                hide obscured y ticks=false,
                minor tick num=1,
                legend style={at={(axis cs:3,-3)}, anchor=north},
                axis lines = middle,
                xlabel=$c$,
                ylabel=utility,
                label style ={at={(ticklabel cs:1.1)}},
                legend cell align={left}]
                    \path[name path=xaxis] (axis cs:0,00) -- (axis cs:4,0);
                    \addplot[very thick,domain=0:1,forget plot]{2*ln(x)};
                    \addplot[name path=e1,very thick, domain=1:3.994]{2*x-2};
                    \addplot[name path=s1, thick,dashed, domain=1:5]{2*ln(x)};
                    \addplot[name path=s2,very thick,forget plot,domain=3.994:5]{2*ln(5*x)};
                    \addplot[name path=s2, thick,dashdotted, domain=0:3.994]{2*ln(5*x)};
                    \addplot [
                            thick,
                            color=black,
                            fill=black, 
                            fill opacity=0.15
                        ]
                            fill between[
                            of=e1 and xaxis,
                            soft clip={domain=1:3.994},];
                    \addlegendentry{\small $v(c,r_G(c))$};
                    \addlegendentry{\small $v(c,0)=f(c,\underline{s})$};
                    \addlegendentry{\small $v(c,1)=f(c,\overline{s})$};
                    \addlegendentry{\small support of $G$};
            \end{axis}
        \end{tikzpicture}
        \centering
    \caption{Example for neutrally stable preferences with hierarchical play: individuals with higher relative consumption play $h$. [$f(c,s) = 2\log(c\cdot s), \; s \in \{1,5\}$]}
    \label{fig:relativeConsHierarchy}
\end{figure}

\begin{figure}[ht!]
        \begin{tikzpicture}[scale=0.75]
        \definecolor{colorG}{RGB}{0,140,90}
        \definecolor{colorR}{RGB}{200,20,70}
            \begin{axis}[grid=none,
                ymin=-8,
                ymax=8,
                xmax=4.5,
                xmin=-0.01,
                samples=101,
                axis x line shift=-0,
                ytick={0},
                xticklabel={},
                yticklabel={},
                hide obscured y ticks=false,
                minor tick num=1,
                legend style={at={(axis cs:3,-3)}, anchor=north},
                axis lines = middle,
                xlabel=$c$,
                ylabel=utility,
                label style ={at={(ticklabel cs:1.1)}},
                legend cell align={left}]
                    \path[name path=xaxis] (axis cs:0,00) -- (axis cs:4,0);
                    \addplot[name path=e1,very thick, domain=0.319:3.994]{0.774+x*0.5};
                    \addplot[thick,domain=0:4,dashed]{2*ln(x)};
                    \addplot[name path=s1,very thick, forget plot, domain=4:5]{2*ln(x)};
                    \addplot[name path=s2,very thick,forget plot,domain=0:0.319]{2*ln(5*x)};
                    \addplot[name path=s2, thick,dashdotted, domain=0.35:5]{2*ln(5*x)};
                    \addplot [
                            thick,
                            color=black,
                            fill=black, 
                            fill opacity=0.15
                        ]
                            fill between[
                            of=e1 and xaxis,
                            soft clip={domain=0.35:4},];
                    \addlegendentry{\small $v(c,r_G(c))$};
                    \addlegendentry{\small $v(c,1)=f(c,\underline{s})$};
                    \addlegendentry{\small $v(c,0)=f(c,\overline{s})$};
                    \addlegendentry{\small support of $G$};
            \end{axis}
        \end{tikzpicture}
        \centering
        \caption{Example for neutrally stable preferences with anti-hierarchical play: individuals with lower relative consumption play $h$.}
    \label{fig:relativeConsAntiHierarchy}
\end{figure}

\noindent\textbf{Labels \& communication.} We assumed that the fitness function is strictly concave in consumption, making any fair gamble costly in terms of expected fitness. A neutrally stable type nevertheless chooses a non-trivial lottery. At least for small gambles, any such implicit fitness cost is outweighed by the benefit from coordinating actions. Lotteries allow individuals to send messages on which actions can be conditioned. Even though in equilibrium not all messages lead to the same fitness outcome, the implicit cost of changing messages prevents stable types from arbitrary switching, excessive risk-taking, as well as bunching on a single message. For a continuous lottery, there are infinitely many such messages. Outcomes that leave individuals with identical messages and don't allow for coordination become a probability $0$ event. This allows for a Pareto-efficient allocation of the social good in the stage game for almost all player pairs.

Such lotteries are, of course, an abstract concept. The fundamental idea, however, is simple. If coordination yields a strong benefit, individuals might be willing to deliberately introduce variability in their observable characteristics. For instance, a person might opt for a riskier investment in the hope to achieve a higher income than their peers, which can be signalled through positional goods. This might then affect their social interactions.%
    \footnote{There is a large body of economic literature on how relative income and conspicuous consumption affect individuals' satisfaction and social interactions (e.g., \citet{clark2009} and \citet{frank05b}). There is also anthropological evidence for the importance of status and its economics consequences. See, for instance, \citet{allen89}, on how prestige affects social interactions among the Enga people.} 
As is shown in this analysis, preferences don't necessarily need to reflect this coordination motive explicitly. If evolution acts on preferences, positional concerns can yield risk-preferences that align evolutionary objectives with individual incentives. Status seeking behaviour can be seen as form of communication with positional goods acting as messages.

This observation closely echoes findings in \citet{herold20}. They analyse a model where individuals can communicate a `label' from a finite message set. In contrast to the setting here, obtaining any such label is free. The messaging stage amounts to cheap-talk. In anti-coordination game, they find essentially two equilibria, which depend on whether the number of available messages are even or odd: (i) a `hierarchical' equilibrium in which the player choosing $h$ is determined by a transitive order on labels. And (ii), an `egalitarian' equilibrium in which each label plays $h$ against some labels and $d$ against the others. These equilibria have a natural analogue in this setting. The utility function described in Section \ref{section:statusPreferenes} yields a hierarchical equilibrium where individuals play $h$ against all opponents with lower consumption levels. Figure \ref{fig:relativeConsHierarchy} illustrates such an equilibrium. Figure \ref{fig:relativeConsAntiHierarchy} visualizes the other extreme, where the individual with the higher relative consumption plays $d$. However, more egalitarian equilibria that lie between the two extremes are also possible. All equilibria have a common characteristic: for all consumption levels, and hence messages, the opponents with different messages can be divided in two sets, one against which an individual plays $h$ and the other against which to play $d$. In equilibrium, these sets contain almost all individuals. For a population of stable types, any such norm that determines which player is to play which action in the stage game renders both expected utility and fitness linear and strictly increasing over any range of randomization.%
    \footnote{This is discussed more formally in Appendix \ref{APPENDIX:addResult}.}
It can deliver both hierarchical and egalitarian outcomes, as departing from any norm that satisfies these criteria cannot be optimal. 

While this shows that there is a clear parallel between costless and costly communication, one could imagine both to occur in conjunction. Depending on the environment, the ways to randomize consumption might be limited. If those only allow for discrete outcomes, there is an additional benefit from adopting costless labels on top of the costly communication. Such an incentive to blend both types of communication can also arise when the perception of costly messages is coarse, meaning individuals might not be able to distinguish all messages from each other. Analysing a setting with such a richer message space and potential constraints to communication might offer an interesting avenue for future research.

\appendix

\section{Proofs}\label{appendix:proofs}

\begin{proof}[\textbf{Proof of Lemma \ref{lemm:coordActionsStage}}]
    Let $G$ be the equilibrium aggregate distribution over messages in the population, given lottery choices $L_{\tau}$ and $L_{\tau'}$ by each type, i.e., $G(c) = (1-\epsilon) \cdot L_{\tau}(c-c_0) + \epsilon \cdot L_{\tau'}(c-c_0)$. Suppose to the contrary that a type $\tau$ is stable against all $\tau' \in \hat{\T}$ and there exists a subset of non-identical message combinations $\hat{M}\times\hat{M}' \subset M\times M$ such that $\phi_{\tau}(m|m') \in (0,1)$ for all $(m,m') \in \hat{M}\times\hat{M}'$, and $G(\hat{M}) \cdot G(\hat{M}')>0$, meaning the set of message combinations has positive measure (according to $G$). Given the interior strategies, in equilibrium either both actions yield the same expected social good, i.e., $S(1, \phi_{\tau}(m'|m)) = S(0, \phi_{\tau}(m'|m))$, or there is a strict best-response, meaning $S(1, \phi_{\tau}(m'|m)) \ne S(0, \phi_{\tau}(m'|m))$. Note that different cases can apply to different subsets of $\hat{M}\times\hat{M}'$. If the latter is true and there is a strict best-response for a (sub-)set $\hat{\underline{M}}\times\hat{\underline{M}}' \subseteq \hat{M}\times\hat{M}'$, $\tau$ cannot be stable against all $\tau' \in \hat{\T}$. To see this, consider a $\tau'$ with preferences over lotteries such that $L$ is chosen for a configuration $(\tau, \tau', 0)$. Then consider preferences over outcomes in the stage game that are identical to $\tau$, except for message combinations $(m,m') \in \hat{\underline{M}}\times\hat{\underline{M}}'$, where there is a strict best-response. In these cases, we can replace the preferences such that a type $\tau'$ chooses the strict best-response. As this type achieves strictly higher fitness than $\tau$ for all message combinations in $\hat{\underline{M}}\times\hat{\underline{M}}'$ and the same fitness for all other message combinations, it follows that $\lim_{\epsilon \to 0} V_{\tau'}(\tau, \tau', \epsilon) - V_{\tau}(\tau, \tau', \epsilon) > 0$. 
    
    Now suppose instead for all $(m,m') \in \hat{M}\times\hat{M}'$, we have $S(1, \phi_{\tau}(m'|m)) = S(0, \phi_{\tau}(m'|m))$, meaning actions $h$ and $d$ yield the same social good allocation in expectation. The anti-coordination nature of the stage game then implies that $\underline{s} > S(0, \phi_{\tau}(m'|m)) = S(1, \phi_{\tau}(m'|m))$. Let $s_{\tau} \coloneqq S(0, \phi_{\tau}(m'|m))$. Consider a type $\tau'$ that plays $\phi_{\tau'}(m|m') = 1$ and $\phi_{\tau'}(m'|m) = 0$ for all such message combinations. This type receives in expectation a social good allocation $s_{\tau'} \ge \epsilon \cdot \underline{s} + (1-\epsilon) \cdot s_{\tau} > s_{\tau}$. As fitness is strictly increasing in $s$, this implies that $\lim_{\epsilon \to 0} V_{\tau'}(\tau, \tau', \epsilon) - V_{\tau}(\tau, \tau', \epsilon) > 0$. It then follows from continuity in payoffs that there exists an $\epsilon' > 0$ such that $V_{\tau'}(\tau, \tau', \epsilon') > V_{\tau}(\tau, \tau', \epsilon')$ for at least some equilibrium, contradicting stability of $\tau$ against $\tau'$.
\end{proof}

\begin{proof}[\textbf{Proof of Lemma \ref{lemm:stageGameStable}}]
    \textbf{Sufficiency:} Let $L \in \Lo$ be the lottery chosen by a type $\tau$ and a type $\tau'$. Let $\theta_{\tau}(m'|m) \in \{0,1\}$ and $\theta_{\tau}(m|m') = 1- \theta_{\tau}(m'|m)$ for almost all $m, m' \in M$. Suppose there is a set $\hat{M} \times \hat{M'} \subset M \times M$ with positive measure (according to $L$) such that $\theta_{\tau'}(m|m') \in (0,1)$. For any $m$, this yields an expected outcome in the stage game for a type $\tau'$ of:
    \begin{equation*}
    \begin{split}
        s_{\tau'}(m) \coloneqq \quad \quad \quad \epsilon& \cdot \bigl[ \theta_{\tau'}(m|m') \bigl(1-\theta_{\tau'}(m'|m)\bigr) \overline{s} + \bigl(1-\theta_{\tau'}(m|m')\bigr) \theta_{\tau'}(m'|m) \underline{s}\bigr]\\
        + (1-\epsilon)& \cdot \bigl[ \theta_{\tau'}(m|m') \bigl(1-\theta_{\tau}(m'|m)\bigr) \overline{s} + \bigl(1-\theta_{\tau'}(m|m')\bigr) \theta_{\tau}(m'|m) \underline{s}\bigr].
    \end{split}
    \end{equation*}
    A type $\tau$ obtains:
    \begin{equation*}
    \begin{split}
        s_{\tau}(m) \coloneqq \quad \quad \quad \epsilon& \cdot \bigl[ \theta_{\tau}(m|m') \bigl(1-\theta_{\tau'}(m'|m)\bigr) \overline{s} + \bigl(1-\theta_{\tau}(m|m')\bigr) \theta_{\tau'}(m'|m) \underline{s}\bigr]\\
        + (1-\epsilon)& \cdot \bigl[ \bigl(1-\theta_{\tau}(m'|m)\bigr) \overline{s} + \theta_{\tau}(m'|m)\underline{s}\bigr],
    \end{split}
    \end{equation*}
    where we made use of the fact that $\theta_{\tau}(m'|m) = 1- \theta_{\tau}(m|m')$.
    Clearly, since $\theta_{\tau'}(m'|m) \in (0,1)$, we have $\lim_{\epsilon \to 0} s_{\tau}(m) > \lim_{\epsilon \to 0} s_{\tau}(m)$. Note further that this extends to $\theta_{\tau'}(m'|m) \in [0,1]$, which given that strategies are non-identical would imply $\theta_{\tau'}(m'|m) = 1-\theta_{\tau}(m'|m)$, which leads to an allocation of $0$ social good in any such pairing. As both types are assumed to choose the same $L$ and as fitness is strictly increasing in $s$, we can conclude that $\lim_{\epsilon \to 0} V_{\tau}(\tau, \tau', \epsilon) - V_{\tau'}(\tau, \tau', \epsilon)>0$ as required.\\
    \\
    \textbf{Necessity:} This follows directly from Lemma \ref{lemm:coordActionsStage}.
\end{proof}

\begin{proof}[\textbf{Proof of Lemma \ref{lemm:lotteryChoice}}]
    Take a population configuration $(\tau, \tau', \epsilon)$ and suppose the resident types preferences over lotteries are such that they choose the trivial lottery in every equilibrium. Consider a sequence of lotteries $\{L_n\}_{n=1}^{\infty}$ with each $L_n$ such that there is a probability mass of $\frac{1}{2n+1}$ at $-\delta$, and a probability mass of $1-\frac{1}{2n+1}$ distributed uniformly over the interval $[0, \frac{\delta}{n}]$, with $\delta < c_0$. Note that every such $L_n$ is a fair and bounded lottery. It follows from Lemma \ref{lemm:stageGameStable} that if a type $\tau$ is stable, then if there are distinct messages, we have $1- \phi_{\tau'}(m'|m) \in \{0,1\}$ and $\phi_{\tau}(m|m') = 1- \phi_{\tau'}(m'|m)$. Suppose $\tau$ follows such a strategy.

    Given the population configuration and lottery choices, the fitness payoff of a type $\tau$ in any equilibrium is bounded above by
    \begin{equation*}
        f_{\tau}(\epsilon) \coloneqq (1-\epsilon) f\bigl(c_0, s_0\bigr) + \epsilon f\bigl(c_0, \overline{s}\bigr),
    \end{equation*}
    where $s_0 \coloneqq \frac{\overline{s}\underline{s}}{\overline{s}+\underline{s}}$, i.e., the expected interior Nash equilibrium payoff, and we made use of Jensen's inequality, noting that $f(c_0, s_0) \ge \E[f(c_0, S(a_i, a_j)) | p_{\text{NE}}(a_i,a_j)]$ for a $f$ weakly concave in $s$, with $p_{\text{NE}}(a_i,a_j)$ the Nash equilibrium probability distribution over action combinations.
    
    The expected fitness of a type $\tau'$ choosing a lottery $L^n$ and stage game strategy $\phi_{\tau'} = \phi_{\tau}$ is bounded below by:
    \begin{equation*}
        f_{\tau'}^n(\epsilon) \coloneqq (1-\frac{1}{2n}) f\bigl(c_0, \underline{s}\bigr)
    \end{equation*}
    where we set the fitness of all types with consumption outcomes $c < c_0$ to 0. Note that $s_0 < \underline{s}$ and hence $f(c_0,\underline{s}) > f(c_0,s_0)$. We can conclude that:
    \begin{equation*}
            \lim_{\epsilon \to 0}\lim_{n \to \infty} f_{\tau}(\epsilon) - f_{\tau'}^n(\epsilon) < 0.
    \end{equation*}
    It follows from continuity that we can find $n^*$ and $\epsilon^*$ such that for every $n>n^*$ and every $\epsilon<\epsilon^*$, we have 
    \begin{equation*}
        V_{\tau'}(\tau,\tau',\epsilon) \ge f_{\tau'}^n(\epsilon)  > f_{\tau}(\epsilon) \ge V_{\tau}(\tau,\tau',\epsilon)
    \end{equation*}
    in some equilibrium if $L^n \in \Lo$. The result follows.
\end{proof}

\begin{proof}[\textbf{Proof of Proposition \ref{propos:linearityLottery}}]
    Take a configuration $(\tau, \tau', \epsilon)$. It follows from Lemma \ref{lemm:coordActionsStage} that if (i) is violated, $\tau$ cannot be stable. Suppose now (i) holds and there is an equilibrium in which type $\tau$ chooses a lottery $L$ and $\tau'$ chooses a lottery $\hat{L} \ne L$, $\; \hat{L} \in \Lo$, while $\phi_{\tau} = \phi_{\tau'}$. Let $\overline{M}_L(c_0)$ be the smallest interval that contains $M_L(c_0)$.\\
   \\
   \textbf{Case} $c_0 \in M_L(c_0)$: Condition (ii) implies that $\int F_{\tau}(m|G) \lambda_L(m-c_0) dm = F_{\tau}(c_0|G)$. Furthermore, (ii) and (iii) together imply that 
   \begin{equation*}
       \int F_{\tau}(m|G) \lambda_{\hat{L}}(m-c_0) dm \le  F_{\tau}(c_0|G) = \int F_{\tau}(m|G) \lambda_{L}(m-c_0) dm.
   \end{equation*}
    Any such lottery$\; \hat{L}$ yields weakly lower expected fitness than $L$ and hence $V_{\tau}(\tau, \tau',\epsilon) \ge V_{\tau'}(\tau, \tau',\epsilon)$.\\
    \\
    \textbf{Case} $c_0 \notin M_L(c_0)$: Take any $c^* \notin M_L(c_0)$ but $c^* \in \overline{M}_L(c_0)$. As any $L$ must be fair, there exist $c_1, c_2, c_3 \in M_L(c_0)$ such that $c_1 < c^* < c_2 < c_3$. Suppose $F_{\tau}(c^*) > \alpha F_{\tau}(c_3) + (1-\alpha) F_{\tau}(c_1)$, with $\alpha = \frac{c^* - c_1}{c_3 - c_1}$. Note that the right-hand side corresponds to the expected (fitness) outcome of an an individual with consumption $c^*$ that chooses a (fair) lottery $L_{\alpha} \in \Lo$ that yields $c_3- c^*$ with probability $\alpha = \frac{c_3 - c^*}{c_3 - c_1}$ and $c_1$ with $1-\alpha$.
    
    Now consider a lottery $L_{\beta}$ that yields $c_3 - c_2$ with probability $\beta = \frac{c_2 - c_1}{c_3 - c_1}$ and $c_1 - c_2$ with $1-\beta$. Clearly, $L_{\beta}$ is also fair and bounded and hence in $\Lo$. It follows that 
    \begin{equation*}
        \int F(m|G)\lambda_{L_{\beta}}(m-c_2) dm = \beta F_{\tau}(c_3|G) + (1-\beta) F_{\tau}(c_1|G) = F_{\tau}(c_2|G)
    \end{equation*}
    where the last equality follows from (ii), noting that $M_{L_{\beta}}(c_2) = \{c_1, c_3\} \subset M_L(c_0)$.
    Consider another lottery $L_{\gamma}$ that yields $c_3 - c_2$ with probability $\gamma = \frac{c_3 - c_2}{c_3-c^*}$ and $c^* - c_2$ with $1-\gamma$. This is again a fair and bounded lottery. By construction: 
    $$
    \gamma F_{\tau}(c_3|G) + (1-\gamma) F_{\tau}(c^*|G) >\gamma F_{\tau}(c_3|G) + (1-\gamma)\bigl[\alpha F_{\tau}(c_3|G) + (1-\alpha) F_{\tau}(c_1|G)\bigr] 
    $$
    But then 
    \begin{equation*}
        \begin{split}
            \gamma F_{\tau}(c_3) + (1-\gamma) F_{\tau}(c^*|G) >& \; \bigl(\gamma + (1-\gamma)\alpha\bigr) F_{\tau}(c_3|G) + (1-\gamma)(1-\alpha) F_{\tau}(c_1|G)\bigr)\\
            =& \; \beta F_{\tau}(c_3|G) + (1-\beta) F_{\tau}(c_1|G),
        \end{split}
    \end{equation*}
    This contradicts (iii). It follows that for any $c^*$ in $\overline{M}_{L}(c_0)$, we have $F_{\tau}(c^*|G) \le \int F_{\tau}(m|G) \lambda_{L^*}(m-c^*) dm$, for any $L^* \in \Lo$ such that $M_{L^*}(c^*) \subseteq M_L(c_0)$. Furthermore, this holds with equality for any $c^* \in M_L(c_0)$. It then follows that 
    $\int F(m|G) \lambda_{L}(m-c_0) dm \ge \int F_{\tau}(m|G) \lambda_{L'}(m-c_0) dm$ for any $L' \in \Lo$ with $M_{L'}(c_0) \subseteq M_{L}(c_0)$. It follows directly from (iii) that this inequality also holds for an arbitrary $\hat{L} \in \Lo$.

    The final step is to show that a departure from a strategy $\phi_{\tau}$ that efficiently correlates actions cannot lead to higher expected fitness if the lottery choice of the resident satisfies (ii) and (iii). 

   If $\phi_{\tau'}$ differs from $\phi_{\tau}$ for a measurable set of message combinations in an equilibrium with an aggregate distribution $G$, then as $\lim \epsilon \to 0$, $F_{\tau'}(m|G)$ approaches
    $$F_{\tau'}(m|G) = \int \Bigl( \phi_{\tau'}(x|m)(1-\phi_{\tau}(m|x))\cdot f(m,\overline{s}) +  (1-\phi_{\tau'}(x|m))(\phi_{\tau}(m|x))\cdot f(m,\underline{s}) \Bigr) dG(x).
    $$
    Using the same argument as in the proof of Lemma \ref{lemm:stageGameStable}, we can conclude that  $\lim_{\epsilon \to 0} F_{\tau'}(m|G) - F_{\tau}(m|G) \le 0$ for all $m \in M_G$, with the inequality strict for any $m$ where $\phi_{\tau'}(m'|m)$ differs from $\phi_{\tau}(m'|m)$ for a set of $m'$ with positive measure according to $G$. Combining this with the previous argument, we can further conclude that for any $\hat{L} \in \Lo$, as $\lim \epsilon \to 0$, we have:
    \begin{equation*}
        \int F_{\tau'}(m|G) \lambda_{\hat{L}}(m-c_0) dm < \int F_{\tau}(m|G) \lambda_{\hat{L}}(m-c_0) dm \le \int F_{\tau}(m|G) \lambda_{L}(m-c_0) dm 
    \end{equation*}
    with $L$ the lottery choice of $\tau$ in equilibrium. Accordingly, $\lim_{\epsilon \to 0} V_{\tau'}(\tau,\tau',\epsilon) - V_{\tau}(\tau,\tau',\epsilon) < 0$. The result follows.
\end{proof}

\begin{proof}[\textbf{Proof of Proposition \ref{propos:statusPreferences}}]
    Let $v: {\R}_+ \times [0,1] \to {\R}_+$ be a Bernoulli utility function over consumption and rank defined as follows:
    \begin{equation}\label{EQ:statusRepres}
        v(c,r) \coloneqq r \cdot f(c,\overline{s}) + (1-r) \cdot f(c,\underline{s}).
    \end{equation}
   Suppose preferences over lotteries in $\Lo$, given an aggregate distribution over messages $G$ and a consumption endowment $c$, can be represented by the following expected utility function:
    \begin{equation}\label{EQ:expectedUtilStat}
        V(L \; | \; c, G) = \int v\bigl(c+x,r_G(c+x)\bigr) \lambda_{L}(x-c) dx,
    \end{equation}
    where $r_G(x)$ captures the measure of individuals with consumption below $x$ (or a convex combination between the measure of individuals strictly and weakly below $x$ in the case of a mass point at $x$), defined as
    \begin{equation*}
        r_G(x) \coloneqq
        \begin{cases}
            G(x),    & \text{if } \lambda_G(x) = 0\\
            \alpha \underline{G}(x) + (1-\alpha) G(x), & \text{otherwise},
        \end{cases}
    \end{equation*}
    for some $\alpha \in (0,1)$ and $\underline{G}(x)$ the left-limit of $G(x)$.
    $V(L \; | \; c, G)$ amounts to the expected utility from consumption and consumption rank outcomes under lottery $L$, given $G$ and $c$. It follows from the proof of Lemma \ref{lemm:lotteryChoice} that a lottery choice from the set of all fair and bounded lotteries that maximizes \eqref{EQ:expectedUtilStat} cannot result in a mass point at any $c>0$ for a pure population in any equilibrium.

    Consider the following strategy in the stage game:
    \begin{equation*}
        \phi(m'|m) = 
        \begin{cases}
        1,    & \text{if } m > m',\\
        0,    & \text{if } m < m',\\
        \frac{f(m,\overline{s}) + f(m,\underline{s})}{f(m,\overline{s}) \cdot f(m,\underline{s})},                  & \text{otherwise}.
    \end{cases}
    \end{equation*}
There exist preferences in the stage game, conditional on $m$ and $m'$, that admit this as a unique equilibrium strategy (i.e., Table \ref{tab:payoff_matrix_rel_stable}). It follows from Lemma \ref{lemm:stageGameStable} that a type with such preferences is evolutionary stable against any other type that has the same choice behaviour over lotteries. The remaining question is whether $V$ leads to equilibrium choices over lotteries that are neutrally stable.

Given Assumptions \ref{AssuCont} and \ref{AssuConcav}, $v$ is continuous and strictly concave in $c$. This means conditions of Proposition 1 (i) of \citet{robray12} are satisfied. This result implies the existence of a unique equilibrium distribution over messages $G$ in a pure population with a corresponding lottery choice $L \in \Lo$ that is among the maximizers of \eqref{EQ:statusRepres}. Furthermore, in any equilibrium, $v(c,r_G(c))$ is linear over the support of $G$ and strictly concave elsewhere. By construction, for a continuous $G$, 
\begin{equation}\label{EQ:eqFandV}
    \begin{split}
        F_{\tau}(m|G)   =& \int \phi_{\tau}(x|m) f(x,\overline{s}) dG(x) + \int \bigl(1-\phi_{\tau}(x|m)\bigr) f(x,\underline{s}) dG(x)\\
                        =& r_G(c) f(m,\overline{s}) + (1-r_G(c)) f(m,\underline{s}).
    \end{split}
\end{equation}
Hence, in any equilibrium, $F_{\tau}(m|G)$ is linear in $m$ over the support of $G$ and strictly concave elsewhere. Conditions (i)-(iii) of Proposition \ref{propos:statusPreferences} are satisfied. Such preferences are neutrally stable.
\end{proof}

\section{Additional Results}\label{APPENDIX:addResult}

\begin{result}
    Suppose there exists a set of types $\hat{\T} \subset \T$ and an $\; \overline{\epsilon}>0$ such that for all distinct $\tau, \tau' \in \hat{T}$ and any configuration $(\tau, \tau', \epsilon)$ with $\epsilon < \overline{\epsilon}$, both types choose an identical continuous lottery $L \in \Lo$ in any equilibrium. Then a type $\tau \in \hat{\T}$ is evolutionary stable against any $\tau' \in \hat{\T}$ if and only if $\phi_{\tau}(m'|m) \in \{0,1\}$ and $\phi_{\tau}(m|m') = 1- \phi_{\tau}(m'|m)$ for almost all $m, m' \in M$ and in any equilibrium there exists a set $\hat{M}\times\hat{M}' \subseteq M \times M$ with positive measure according to $L$ such that $\phi_{\tau'}(m'|m) \in (0,1)$ for any $(m',m) \in \hat{M}\times\hat{M}'$. 
\end{result}

\begin{proof}
    This follows directly from the proof of Lemma \ref{lemm:stageGameStable}.
\end{proof}

Proposition \ref{propos:linearityLottery} derives sufficient conditions for stability. Linearity over the support of $G$ (i.e., (ii)) is sufficient but not necessary for stability for every $\epsilon < \overline{\epsilon}$, as strategies can change discontinuously at $\epsilon = 0$. Imposing a notion of continuity allows a strengthening of the result.

\begin{define}[Continuous strategies] We say strategies are \textbf{continuous} if for any sequence of population configurations $\bigl\{ (\tau, \tau', \epsilon^n) \bigr\}^{\infty}$, with $\bigl\{\epsilon^n\bigr\}^{\infty}$ monotonically converging to $0$, the corresponding equilibrium distributions $\bigl\{G^n\bigr\}^{\infty}$ converge to $G$ in the weak topology and $\lim_{n \to \infty} F_\tau(m|G^n) = F_\tau(m|G)$. 
\end{define}

Observe that for preferences defined as in \eqref{EQ:statusRepres}, this is satisfied. The Bernoulli utility $v$ is a continuous function. By definition of convergence in the weak topology, $v(c,r_{G^n}(c)) \to v(c,r_{G}(c))$, as $G^n \to G$. It then follows from \eqref{EQ:eqFandV} that the same applies for the corresponding $F_{\tau}(m|G^n)$. Accordingly, preferences over lotteries approach $V(L|c_0,G)$, as $\epsilon \to 0$. This ensures that indeed $G^n \to G$. With such strategies, linearity as well as monotonicity over the support of $G$ become necessary conditions. 

\begin{result}\label{result:linearityAndMonotonicity} If strategies are continuous in the distribution over messages, then a type $\tau \in \T$ is neutrally stable only if for any equilibrium of the pure population with distribution $G$, $F_{\tau}(m|G)$ is strictly increasing and linear in $m$, for almost all $m$ in the support of $G$.
\end{result}

\begin{proof}
    Let $L_{\tau}$ be the lottery choice corresponding to such a $G$.
    Suppose not and there is an interval $I \subset \supp(L_{\tau})$ such that $F_{\tau}(m|G)$ is weakly decreasing in $m$ over $I$ and strictly increasing elsewhere. 
    Note that for a stability, $L_{\tau}$ needs to be continuous. Then for almost all $m \in \R_+$, $F_{\tau}(m|G)$ is bounded below by 
     $f(m,\underline{s})$, which is strictly concave in $m$. This implies that $F_{\tau}(m|G)$ can only be weakly decreasing over a bounded interval. Let $\overline{m} \equiv \sup I$, which by the previous argument must exist. Then for almost all $m \in I$, any fair gamble $(1-\alpha) (m-\epsilon) + \alpha \overline{m}+\epsilon$ must deliver expected fitness strictly greater $F_{\tau}(m|G)$ for an arbitrarily small $\epsilon$ and the $\alpha \in (0,1)$ that makes the lottery fair. In fact, we can find a fair and continuous lottery that delivers a strict benefit. It can be easily verified that the same is true if $I$ is increasing but strictly convex over an interval $I$.
     
     This implies that there exists a mutant type $\tau'$, with preferences that lead to the choice of such a beneficial gamble,  that achieves strictly higher expected fitness. Let $G^{\epsilon}$ be the equilibrium distribution for the population distribution $(\tau,\tau', \epsilon)$. Clearly, as $\epsilon \to 0$, we have $G^{\epsilon} \to G$. As strategies are required to be continuous in the distribution over messages, a beneficial gamble must exist for any $G^{\epsilon}$, with $\epsilon$ sufficiently small. It follows that $\tau$ is not stable.
\end{proof}

\newpage

\bibliographystyle{plainnat}
\bibliography{status_resource_bib}

\end{document}